\documentclass[12pt]{article}

\usepackage{graphicx}				
\usepackage{amsmath}
\usepackage{amsfonts}

\DeclareMathOperator*{\argmin}{arg\,min}

\usepackage{xspace}
\def\osprey{{\sc{osprey}}\xspace}
\def\bwmstar{BWM$^*$}

\def\bbks{\textit{BBK$^*$}\xspace}

\title{Protein Design by Algorithm}
\author{Mark A. Hallen$^1$ and Bruce R. Donald$^2$}

\begin{document}
\maketitle
$^1$ Toyota Technological Institute at Chicago, Chicago, IL 60637

$^2$ Duke University, Durham, NC 27708



\section{Abstract}
We review algorithms for protein design in general. Although these algorithms have a rich combinatorial, geometric, and mathematical structure, they are almost never covered in computer science classes. Furthermore, many of these algorithms admit provable guarantees of accuracy, soundness, complexity, completeness, optimality, and approximation bounds.  The algorithms represent a delicate and beautiful balance between discrete and continuous computation and modeling, analogous to that which is seen in robotics, computational geometry, and other fields in computational science.  Finally, computer scientists may be unaware of the almost direct impact of these algorithms for predicting and introducing molecular therapies that have gone in a short time from mathematics to algorithms to software to predictions to preclinical testing to clinical trials.  Indeed, the overarching goal of these algorithms is to enable the development of new therapeutics that might be impossible or too expensive to discover using experimental methods.  Thus the potential impact of these algorithms on individual, community, and global health has the potential to be quite significant. 

\section{The problem of computational protein design}\label{cpd_problem}

Proteins are a class of large molecules that are involved in the vast majority of biological functions, from cell replication to photosynthesis to cognition.  The chemical structure of proteins is very systematic~\cite{alg_smb_textbook}---they consist of a chain of atoms known as the~\textit{backbone}, which consists of three-atom (nitrogen-carbon-carbon) repeats known as~\textit{residues}, each of which features a ~\textit{sidechain} of atoms emanating from the first carbon.  There are in general 20 different options for sidechains, and a residue with a particular type of sidechain is known as an~\textit{amino acid} (so there are also 20 different amino acid types).  For billions of years, the process of evolution has optimized the sequence of amino acids that make up naturally occurring proteins to suit the needs of the organisms that make them.  So we ask: can we use computation to design non-naturally-occuring proteins that suit our biomedical and industrial needs?  

This question is a combinatorial optimization problem, because the output of a protein design computation is a sequence of amino acids.  Due to the vast diversity of naturally occurring proteins, it is possible---and very useful---to begin a protein design computation with a naturally occurring protein and then to modify it to achieve the desired function.  In this review, we will focus on protein design algorithms that perform this optimization using detailed modeling of the 3-D structure of the protein~\cite{alg_smb_textbook,cosb_design}.  Thus they will begin with a~\textit{starting structure}, a 3-D structure of a (typically naturally occurring) protein that we wish to modify.  

To illustrate this with an example, imagine we wish to perform a simple example modification to a protein to make it more stable, so it can still function at higher temperatures.  In this case we must minimize the protein's energy with respect to its sequence of amino acids.  In structure-based design, energy is typically estimated using~\textit{energy functions}, which map the 3-D geometry of a molecule to its energy, so the optimization becomes slightly more complex: we minimize the energy with respect to both the~\textit{sequence} (of amino acids) and the~\textit{conformation} (the 3-D geometry of the protein, i.e., the locations of all its atoms in space).  While the sequence is a discrete variable, the conformation is a continuous one because coordinates in $\mathbb{R}^3$ are continuous variables.  There are some physical (e.g., holonomic) constraints on how atoms can move relative to each other, and thus the conformational space can most effectively be represented using internal coordinates, resulting in the joint angle {\em configuration space} familiar in robotics and motion planning in computer science.  Nevertheless, the full conformational space of a protein is too vast to search exhaustively, especially with a simultaneous search over sequence space.  

Computational structure-based protein design arose as a response to this difficulty.  Its initial goal was to overcome certain combinatorial obstructions to designing with a discretized version of the conformational space. Hence, in order to study protein design, it is first necessary to understand the structure of this simpler (but still non-trivial) discrete optimization problem. To this end, we first give a flavor for the issues that arise in discrete optimization.  We examine a very special case---the case of discrete rotamers  and a simple Markov random field (MRF)-like energy function (Section~\ref{discrete_pairwise}).  Next, we carefully define a mixed discrete-continuous optimization problem that gives sidechains and then backbones continuous flexibility within a conformational voxel (Section~\ref{cont_flex_section}).  Then, we present algorithms that compute partition functions over many states, analogously to well-known statistical inference and machine learning computations (Section~\ref{free_energy}), and that exploit improved, more realistic energy functions (Section~\ref{improved_efunc}).  

It is also often useful in protein design to optimize objectives other than simply the energy of a protein.  However, many useful design objectives can still often be posed in terms of the energies of multiple~\textit{biophysical states} of a protein---for example, states where it is bound to particular other molecules.  Thus, the problem of~\textit{multistate design}, which we will formalize in Section~\ref{multistate_problem}, is appropriate for tasks like optimizing the binding of one protein to another molecule, or even specific binding to a second molecule while excluding binding to a third molecule.  Together with some novel types of objective functions, discussed in Section~\ref{exotic_objfcn}, multistate design is a more general tool to optimize the desired function of a protein with respect to sequence.  

We will highlight the computational techniques employed for each of these problems.  These include techniques from combinatorial optimization, constraint satisfaction, machine learning, and other areas.  For the relatively simple protein design problems addressed in this review, we find that algorithms with a beautiful mathematical structure suffice.  This permits us to illustrate by specific examples the situation confronting practical protein designers in academic or biopharmaceutical laboratories.  Throughout, we review algorithms that are of intrinsic mathematical interest and with the potential for high impact on the engineering of new molecular therapies for human disease. 

In addition to this review of core algorithmic work, we will briefly discuss methods to accelerate protein design computations using GPU hardware (Section~\ref{gpu}), as well as some cases in which computationally designed proteins have performed well in experimental tests (Section~\ref{applications}).  Protein design has already had success in the design of novel enzymes, proteins with non-naturally-occurring structures, and proteins with therapeutic applications.  As the field matures we expect to see even more successes from this promising technique.  

\section{The pairwise discrete model}\label{discrete_pairwise}

\subsection{Problem definition}\label{discrete_pairwise_problem}

We will now formalize this problem of stabilizing a protein, using some simplifying assumptions, which will yield the most commonly used mathematical formulation of the protein design problem.  This review will present several algorithms to attack this problem, and also enhancements to the formulation with more sophisticated objectives and/or modeling assumptions.  

Changing the sequence of a protein---i.e.,~\textit{mutating} it---does not alter the chemical structure of its backbone\footnote{Actually there is one amino acid, proline, whose sidechain bonds to the backbone in two places, but it does not alter the repeating nitrogen-carbon-carbon pattern of backbone atoms.  }, and the largest conformational changes are typically found in sidechains near the site of the mutations (we will designate these residues as~\textit{flexible}, i.e., we will consider it necessary to search their conformational space).  Thus, we will assume that the backbone conformation (and possibly some of the sidechain conformations, for residues farther from the site of mutations) is the same as in the starting structure.  Moreover, analyses of sidechain conformational space have found sidechain conformations for each amino-acid type to occur in clusters known as~\textit{rotamers}.  We will refer to the modal sidechain conformation in each cluster as an~\textit{ideal rotamer}.  Then, for the sidechains with respect to whose amino-acid type and conformation we wish to optimize, we will assume that the sidechain conformations will be ideal rotamers, meaning we need only optimize over a discrete set of (sequence, conformation) pairs in which each residue must be assigned an amino-acid type and one of the ideal rotamers for that amino-acid type.  

Let $\mathbf{r}$ be a list of rotamers (which may be of any amino-acid type) for the residues that we are treating as flexible and/or mutable.  If we use only ideal rotamers, $\mathbf{r}$ fully defines a sequence and conformation for the protein, so our energy function gives us a well-defined energy $E(\mathbf{r})$, and our optimization problem becomes simply finding $\argmin\limits_\mathbf{r} E(\mathbf{r})$.  However, one more simplifying assumption is often applied: that we are using a~\textit{pairwise energy function}, which is a sum of terms that each depend on the amino-acid types and conformations of at most two residues.  In this case, we can expand
\begin{gather}
E(\mathbf{r}) = \sum_i E(i_\mathbf{r}) + \sum_{j<i} E(i_\mathbf{r},j_\mathbf{r})
\label{pairwise_E}
\end{gather}
where $i$ and $j$ are residues, and $i_\mathbf{r}$ is the rotamer that $\mathbf{r}$ assigns to residue $i$ (we place the residue position in the subscript, following the convention of the field).  The pairwise energy function gives us a well-defined 1-body energy $E(i_r)$ and 2-body energy $E(i_r,j_s)$ for any rotamers $i_r$ and $j_s$, and indeed these energies can be precomputed (generating an~\textit{energy matrix}) before the process of optimization begins, allowing the optimization to simply operate on the energy matrix rather than calling the energy function directly.  Thus, we can formalize the protein design problem in this simple pairwise discrete model as
\begin{gather}
\argmin_\mathbf{r} \left( \sum_i E(i_\mathbf{r}) + \sum_{j<i} E(i_\mathbf{r},j_\mathbf{r}) \right).
\label{pairwise_discrete_cpd}
\end{gather}
We will refer to the solution of Eq.~\eqref{pairwise_discrete_cpd} as the~\textit{global minimum-energy conformation}, or GMEC.  This problem is equivalent to finding the maximum-likelihood solution for a Markov random field with only pairwise couplings~\cite{alg_smb_textbook,BPMultispecific}.  

Finding the GMEC is unfortunately NP-hard even to approximate~\cite[building on a result of Winfree]{NP-hard_approx_GMEC}.  But much algorithmic and development work has attacked it, and most biophysically relevant cases of the problem can be solved efficiently in practice with provable guarantees of accuracy.  We now review some of this work.   

Work on this problem using heuristic protocols such as simulated annealing, Monte Carlo simulation, and genetic algorithms is surveyed comprehensively in~\cite{alg_smb_textbook,cosb_design}.  Moreover, Monte Carlo simulation in this context is often not ergodic, rendering it less reliable than mathematical methods like Monte Carlo integration that can obtain accurate error bars based on the variance of an ergodic simulation.  As a result, estimates of the GMEC even from a highly optimized Monte Carlo/simulated annealing protocol exhibit empirically significant deviations from the true optimum~\cite{Simoncini}.  Similar empirical results have been found in several other areas of structural biology requiring global minimizers, as reviewed in~\cite{cosb_design}.  For these reasons, in this review we concentrate on provable algorithms that may be of greater interest to computer scientists.  

\subsection{Approaches to the problem}
\subsection{The classic DEE/A* framework}
The first breakthrough toward solving Eq.~\eqref{pairwise_discrete_cpd} was the DEE algorithm~\cite{DEE} (with refinements due to Goldstein), which eliminates rotamers that cannot be part of the GMEC.  It works by comparing two rotamers $i_r$ and $i_t$ for the same residue.  $i_r$ can be pruned if every conformation $\mathbf{r}$ containing $i_r$ is higher in energy than the corresponding conformation in which $i_r$ has been replaced by $i_t$, i.e., if
\begin{gather}
\min_\mathbf{r} \left( E(i_r) - E(i_t) + \sum_{j\neq i} E(i_r,j_\mathbf{r}) - E(i_t,j_\mathbf{r}) \right) > 0.
\label{DEE_unrelaxed}
\end{gather}
Evaluating Eq.~\eqref{DEE_unrelaxed} is as hard as finding the GMEC directly.  But the sum of minima is always a lower bound for the minimum of a sum, so we obtain the following sufficient condition for Eq.~\eqref{DEE_unrelaxed}, which can be evaluated in time linear in the number of residues:
\begin{gather}
E(i_r) - E(i_t) + \sum_{j\neq i} \min_s \bigg( E(i_r,j_s) - E(i_t,j_s) \bigg) > 0.
\label{DEE}
\end{gather}
We call Eq.~\eqref{DEE} the~\textit{DEE criterion}.  By evaluating it for each residue $i$ and each pair of rotamers $i_r$ and $i_t$ that are available at $i$, we can greatly prune the space of rotamers that may be part of the GMEC.  This pruning step is polynomial-time~\cite{alg_smb_textbook}.  Thus, the combinatorial bottleneck must occur later, in the enumeration step.  We describe this below.    

DEE is an efficient algorithm, but it still may leave multiple possible rotamers for some or all of the residues.  This problem has been solved by deploying the A* algorithm from artificial intelligence to find the GMEC using only the rotamers remaining, i.e., using DEE/A*~\cite{DEE/A*}.  Briefly, the A* algorithm in this context builds a priority queue of nodes that represent a partially defined conformation $\mathbf{q}$, which consists of rotamer assignments for only a subset $S(\mathbf{q})$ of the residues.  The~\textit{score} of a node is a lower bound on the energy of any conformation containing all the rotamers in $\mathbf{q}$ (i.e., on $\min\limits_{i_\mathbf{r}=i_\mathbf{q}\,\forall\,i\in S(\mathbf{q})} E(\mathbf{r})$).  We repeatedly extract the lowest-scoring node from the queue and expand it by creating nodes for which one more residue has a defined rotamer. Eventually the lowest-scoring node will be a fully-defined conformation.  Since all conformations in other nodes must have higher energies (based on the nodes' lower bounds), this fully-defined conformation must be the GMEC.  

This shows that it is possible to find the GMEC with guaranteed accuracy, and indeed to do so significantly faster (in practice) than exhaustive enumeration of conformations.  We will now discuss even more sophisticated and efficient algorithms for this problem.  

\subsection{Algorithms from weighted constraint-satisfaction problems}
One source of such improved algorithms is from the field of weighted constraint-satisfaction problems (WCSPs), of which the pairwise discrete protein design problem (Eq.~\ref{pairwise_discrete_cpd}) can be seen as a special case.  To use these techniques, the energy matrix is encoded as a cost-function network (CFN), which includes the same type of 1- and 2-body terms as an energy matrix from protein design~\cite{wcsp_jcc}. The most efficient provably accurate algorithms for WCSPs perform a tree search like A*, but with much more refined heuristics to guide the search (including both upper and lower bounds).  They also usually employ a depth-first branch-and-bound approach rather than a best-first search like A*.  As a result, far less memory is required in practice.  A large set of empirical benchmarks in~\cite{CFN_design} showed that the Toulbar package for WCSPs significantly improved the state-of-the art efficiency for protein design in the discrete pairwise model.  Moreover, this increase in efficiency allowed direct comparison of the true GMEC (computed by WCSP algorithms) to estimated GMECs from the popular but non-provable simulated annealing algorithm, as implemented in the Rosetta software, for very large protein design problems.  Significant discrepancies were found~\cite{Simoncini}, and indeed the error in simulated annealing's estimates increased with protein size.  This highlights the need for algorithms with provable guarantees for protein design, as described in this work.  

A related and also provable approach is to reduce Eq.~\eqref{pairwise_discrete_cpd} to an integer linear programming problem~\cite{ILP_design}.  

\subsection{Algorithms making sparsity assumptions}
\begin{figure}
\includegraphics[width=3.3in]{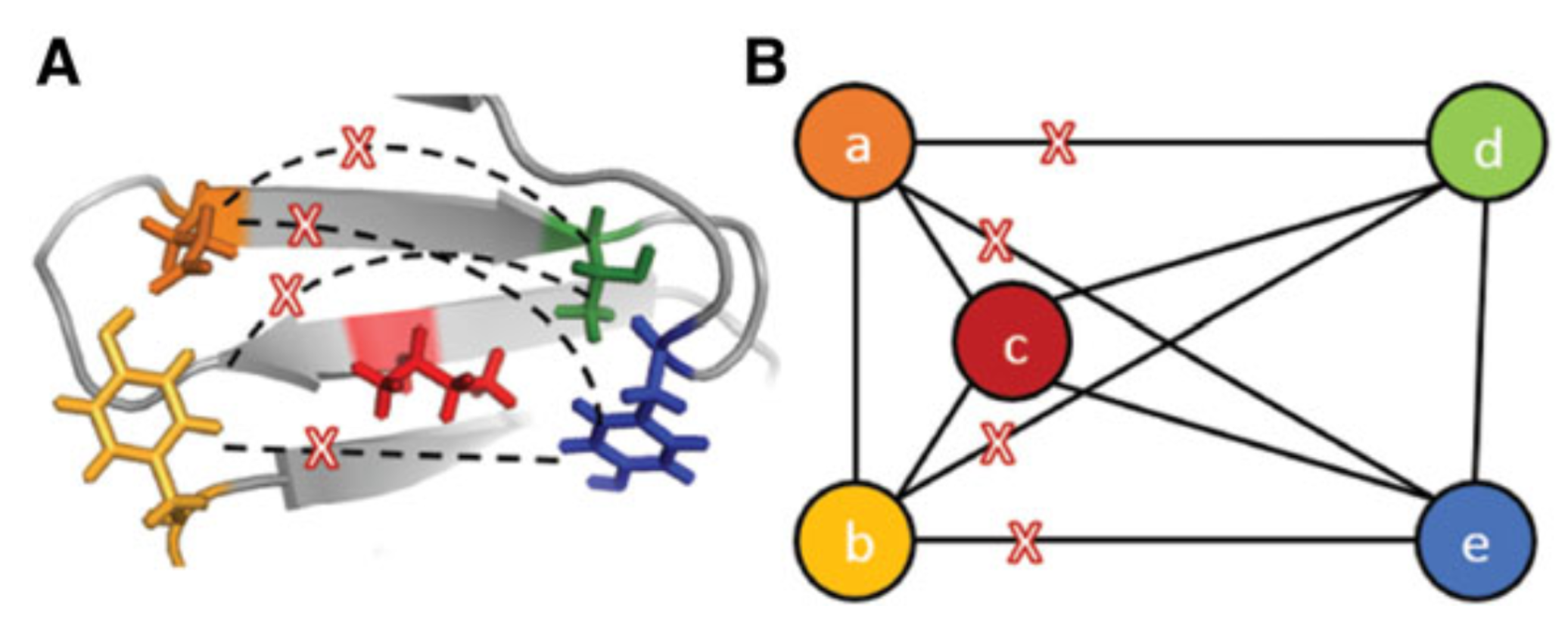}
\caption{(A) Pairwise energy functions compute energies between pairs of mutable residues (colored) in a protein design problem, but in practice many pairs have very small interaction energies (marked with X's).  (B) A sparse residue interaction graph (SPRIG) has mutable residues as nodes; edges with small interaction energies can be deleted, enabling highly efficient protein design computations.  Figure adapted with permission from~\cite{BWM*}.}  
\label{sprig}
\end{figure}

Although protein design as expressed in Eq.~\eqref{pairwise_discrete_cpd} is NP-hard even to approximate~\cite{NP-hard_approx_GMEC}, it is possible to add additional assumptions that make it solvable in polynomial time.  Suppose we assume that some pairs of residues have uniformly zero interaction energies, such that the graph whose nodes are residues and whose edges denote residue pairs with nonzero 2-body energies is sparse, making it a~\textit{sparse residue interaction graph (SPRIG)} (Fig.~\ref{sprig}).   The TreePack algorithm~\cite{treewidth} can find the GMEC in polynomial time when the SPRIG has constant tree-width.  Moreover, the \bwmstar~algorithm can find the GMEC in polynomial time and also efficiently enumerate the $k$ best conformations in gap-free order when the SPRIG has constant branchwidth (where $k$ is requested by the user).  

\section{Improved models}\label{improved_models}

The pairwise discrete model (Eq.~\ref{pairwise_discrete_cpd}) captures the most essential aspects of computational protein design, but it falls short for many practical applications.  Despite the prevalence of rotameric conformations of protein sidechains, real proteins do have significant {\em continuous flexibility} in the neighborhood of each ideal rotamer.  Backbone motions due to mutations are often non-negligible as well.  Moreover, the energy model in Eq.~\eqref{pairwise_discrete_cpd} falls short in two ways: the most accurate energy functions are not explicitly pairwise, and the behavior of a protein is actually determined by its~\textit{free energy}---a quantity based on the distribution of its conformations' energies---rather than on the single minimum-energy conformation.  Finally, as mentioned in Section~\ref{cpd_problem}, it is often useful to have a more sophisticated objective function than simply minimizing the energy of a single biophysical state of a protein.  In this section, we will review algorithms to address these five shortcomings ({\em vide supra}) of the discrete pairwise model of protein design.  

\subsection{Continuous flexibility}\label{cont_flex_section}

\subsubsection{Defining the problem} The problem of continuously flexible protein design differs from Eq.~\eqref{pairwise_discrete_cpd} in that each rotamer is no longer a single conformation of its residue.  Rather, each rotamer is a set of conformations, which we can model as a~\textit{voxel} in the form of bounds on each of several continuous internal coordinates.  Sidechain flexibility in proteins occurs mainly in the form of changes in dihedral angles, and thus the conformation space of a protein can be modeled accurately as a union of voxels in dihedral angle space.  For example, in~\cite{minDEE}, each voxel is centered at an ideal rotamer, and allows up to $\pm9^\circ$ of flexibility in each dihedral angle in either direction from the ideal rotamer's dihedral angle.  The problem is then to find the list of rotamer assignments $\mathbf{r}$ whose voxel contains the lowest-energy conformation---the~\textit{minGMEC}.  

\begin{figure}
\includegraphics[width=1.6in]{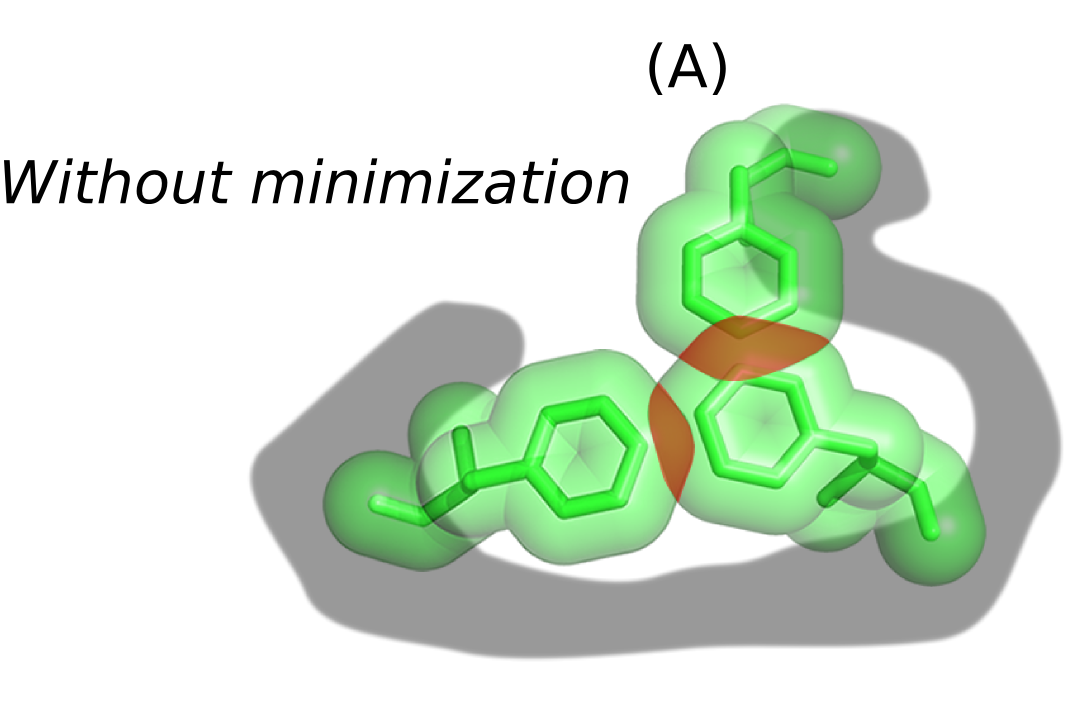}
\includegraphics[width=1.6in]{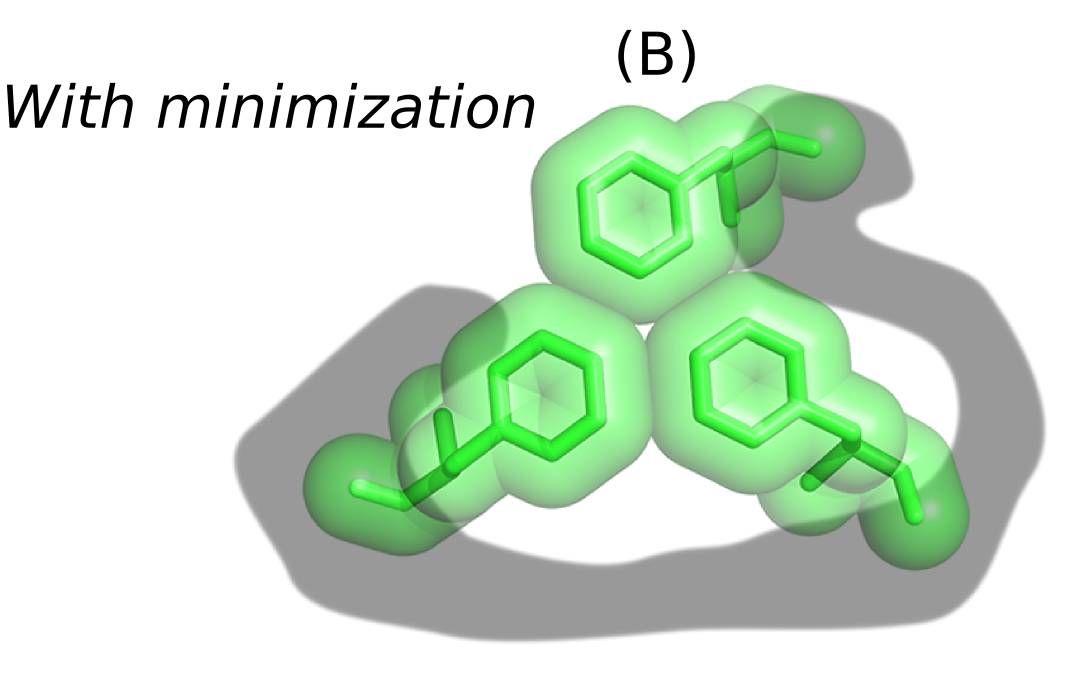}
\caption{(A) A conformation modeled using ideal rotamers may have steric clashes---atom pairs that are unphysically close together---even when (B) continuous minimization of the conformation's energy, without changing the rotamers of any residues, results in a very favorable energy.  This underscores the need to account for continuous flexibility throughout sequence and conformational search for protein design.  Figure adapted with permission from~\cite{iMinDEE}. }
\label{cont_flex}
\end{figure}

This problem has both discrete and continuous components, much like AI planning, where there are discrete steps like STRIPS or TWEAK and continuous steps like motion planning.  Like robust optimization, its aim is to prevent error due to insufficiently fine sampling of conformational space---we wish to avoid eliminating a rotamer merely because its ideal rotameric conformation appears unfavorable, since a small continuous adjustment may turn out to make it optimal.  Indeed, it is relatively common for ideal rotamers to be physically infeasible due to a clash (a pair of atoms too close to each other), but for a small continuous adjustment to suffice to find a favorable conformation~\cite{minDEE,iMinDEE,OSPREY_MIE} (Fig.~\ref{cont_flex}).  Moreover, the optimal sequence is often significantly different, and more biophysically realistic, when continuous flexibility is taken into account than when it is neglected~\cite{iMinDEE,OSPREY_MIE}.  

Notably, no benefit in design is obtained by simply performing a discrete optimization and then continuously minimizing the energy of the discrete GMEC~\textit{post hoc}: such minimization does not change the optimal sequence that is selected.  Rather, to obtain the full benefits of continuous flexibility, one must perform~\textit{minimization-aware} design that finds the minGMEC with guarantees of accuracy by taking continuously flexibility into account from the beginning.  There are two general approaches to minimization awareness.  

\subsubsection{Adapting discrete algorithms to bound the continuous problem}

Algorithms for discrete protein design can be adapted to be minimization-aware by having them prune using~\textit{bounds} on conformational energies, rather than on conformational energies directly.  If a list of voxels $\mathbf{r}$ represents a region in conformational space rather than a single conformation, then its energy (Eq.~\ref{pairwise_E}) may not be well defined~\textit{per se}, but a lower bound on its energy can be expressed in the form of Eq.~\eqref{pairwise_E}, simply by minimizing each of the 1- and 2-body energy terms over the voxel.  Discrete protein design algorithms can then be used to enumerate conformations in order of lower bound.  Once these conformations have been continuously minimized, additional conformations can be pruned based on their lower bounds as well, allowing provable computation of the minGMEC.  This approach has been developed effectively by~\cite{minDEE,iMinDEE}, who adapt the entire DEE/A* framework to be minimization-aware.    

Other discrete algorithms also fit well into the framework of minimization-awareness based on bounds.  For example, both belief propagation (BP) and the self-consistent mean field method (SCMF) are usually employed to estimate a GMEC, with no proofs of closeness to the optimal solution.  However, SCMF can generate a provably correct lower bound on the GMEC energy, while tree-weighted belief propagation can generate a provably correct upper bound.  Thus, by operating on bounds, both algorithms become provable.  This contrasts with the exact rigid energies used with methods from weighted constraint satisfaction and integer linear programming.  

\subsubsection{Reducing the continuous problem to a discrete one}

A more recent approach to minimization-aware protein design is based on machine learning and reducing the continuous protein design problem to a discrete one, without significantly compromising accuracy.  Although the energy of a voxel $\mathbf{r}$ is not explicitly in the form required for discrete protein design algorithms (Eq.~\ref{pairwise_E}), there is a well-defined energy $E(\mathbf{r})$ (generally the continuously minimized energy) that we want to optimize, and we can fit it to the form of Eq.~\eqref{pairwise_E} using machine learning.  This approach is very efficient, as implemented in the LUTE algorithm~\cite{LUTE_RECOMB}, and also accommodates other improvements in biophysical modeling\footnote{Such as non-pairwise energy functions, including those modeling solvation effects (see Section~\ref{improved_efunc}), quantum chemistry, and continuous entropy.}, because the user can choose the function $E(\mathbf{r})$ that is taken as input.  The implementation of LUTE described in~\cite{LUTE_RECOMB} also incorporates some elements of the bound-based approach to continuous flexibility, because it uses iMinDEE~\cite{iMinDEE}, a minimization-aware version of DEE, as a preprocessing step, resulting in a critical improvement in its training and test error.    

\subsubsection{Backbone flexibility}

\begin{figure}
\includegraphics[width=1.6in]{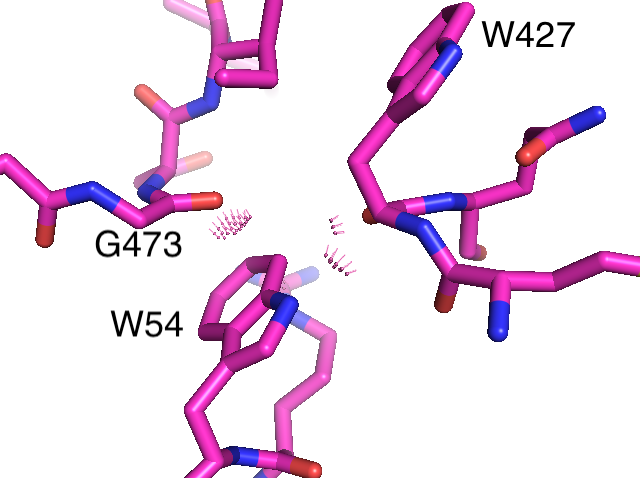}
\includegraphics[width=1.6in]{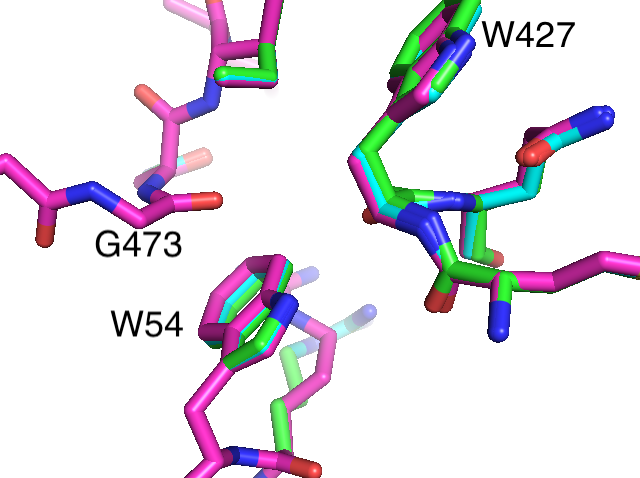}
\caption{Left: Mutating residue 54 of the anti-HIV antibody VRC07 to the amino-acid tryptophan (W) improves its function in experimental tests~\cite{VRC07_enhance}, but rigid-backbone modeling of this mutation shows unavoidable steric clashes (purple conformation).  Right: CATS finds a non-clashing conformation (green), resolving this conundrum, while DEEPer (blue) alleviates the clashes partially.  Figure adapted with permission from~\cite{CATS}.  }
\label{CATS}
\end{figure}

Continuous sidechain flexibility handles discrepancies between ideal rotamers and the actual sidechain conformation.  But an additional type of continuous flexibility---backbone flexibility---is necessary to handle discrepancies between the starting structure's backbone conformation (experimentally observed for the original sequence) and the backbone conformation that is optimal for each mutant sequence.  Like continuous sidechain flexibility, backbone flexibility can be handled using voxels, which can bound the backbone's continuous internal coordinates in a neighborhood around the starting structure's backbone.  The main difference is that the choice of internal coordinates is less straightforward---one must find coordinates that adequately represent the biophysically important backbone flexibility in the vicinity of the mutations without obtaining an intractably large conformational space to search.  These are properties that are satisfied by~\textit{sidechain} dihedrals, whose locality makes them the obvious choice of internal coordinates for sidechains. But they are not satisfied by the standard~\textit{backbone} dihedrals $\phi$ and $\psi$, because local changes in the backbone dihedrals will propagate throughout the protein, disrupting its large-scale structure unless the changes are very small.  The DEEPer algorithm~\cite{DEEPer} addresses this problem by using only backbone motions based on experimental observations, such as the backrub motion observed in crystallographic alternates.  The CATS algorithm~\cite{CATS} allows a larger degree of continuous motion by constructing a new type of backbone internal coordinates that can model the local motion of a contiguous segment of the protein backbone in all biophysically feasible directions (Fig.~\ref{CATS}).  Both algorithms can be used in conjunction with continuous sidechain flexibility modeling and design.  

\subsection{Multistate design}
\subsubsection{Defining the multistate problem}\label{multistate_problem} 
Protein design software is already quite effective at stabilizing proteins, but we must pursue other objectives if it is truly to meet the full range of biomedical and bioengineering needs for modified proteins.  Most of the important objectives involve binding---for example, binding to a protein in the human body that is involved with disease, and also not binding to other, possibly similar, proteins that are essential to normal functioning of the body.  These objectives can be modeled in terms of multiple~\textit{biophysical states}---states in which the protein being designed is unbound, bound to a particular desired target, or bound to a particular undesired target, etc.  Each state $a$ has an energy $E_a(s)$, which we can approximate as the energy of the lowest-energy conformation for the state (as a function of sequence $s$).  We want favored states to be low in energy and unfavored states to be high in energy, since this will cause the protein to adopt the favored states in preference to the unfavored ones.  

Thus, following~\cite{COMETS}, we can pose the problem of~\textit{multistate design} as a kind of linear programming on protein state energies.  We will define~\textit{linear multistate energies (LMEs)}, which are functions of sequence $s$, in the form
\begin{gather}
c_0 + \sum_a c_a E_a(s),
\end{gather}
where the coefficients $c$ are chosen by the user.  For example, to make an LME representing the binding energy between the protein we're designing and another molecule, we would set $c_b=1$ and $c_u=-1$ where $b$ is the bound state and $u$ is the unbound state.  We then wish to minimize not a single state's energy, but an LME, with respect to sequence.  We may also wish to constrain other LMEs to have values above or below a user-specified threshold---for example, we may wish to keep the binding energy to a undesired target higher than the observed binding energy of the unmutated protein to that undesired target.   

\subsubsection{Algorithms for multistate design} 

The formulation in Section~\ref{multistate_problem} comes from~\cite{COMETS}, who also present the first provable algorithm to solve this problem without exhaustive enumeration of sequences.  This algorithm, COMETS, builds an A* tree with nodes representing partial sequences.  Conformational search is handled with a combination of bounding techniques and construction of a ``tree within a tree'' for each promising sequence.  The main tree is thus responsible for sequence search, while the inner trees each correspond to a single node of the main tree and perform conformational search for the sequence corresponding to that node.  

DEE itself has also been adapted for multistate design.  Specifically, within each sequence and biophysical state, multistate design (as defined in Section~\ref{multistate_problem}) is simply computing a GMEC, and as a result it is provably accurate to perform DEE pruning within each biophysical state as long as only competitor and candidate rotamers of the same amino-acid type are considered~\cite{typedep}.  This technique is known as~\textit{type-dependent DEE}.  The multistate design problem has also been addressed using belief propagation~\cite{BPMultispecific} and with a variant of simulated annealing (albeit without any guarantees of accuracy)~\cite{RECON}.  

As in the case of continuous flexibility, machine learning has yielded a novel and very promising technique for multistate design.  The {\em cluster expansion} technique calculates energies for a training set of sequences (for each state) and then learns an energy function that is a sum of terms dependent only on 1 or a few residues' amino acid types.  In this formulation, multistate design becomes mathematically equivalent to discrete single-state design, although combinatorially easier because there are fewer amino acid types than possible rotamers.  This technique has yielded designer peptides with high selectivity for their desired target in experimental tests~\cite{CLASSY}.  

Finally, other formulations of multistate design besides that in Section~\ref{multistate_problem} have been used quite fruitfully.  The paradigm of meta-multistate design~\cite{MetaMultistateDesign}, which accounts for protein dynamics, has yielded designed proteins known as DANCERS (Dynamic and Native Conformational ExchangeRs), which not only exchange between specified conformational states, but do so on the timescale of milliseconds.  

\subsection{Improved energy modeling}
We have so far taken the energy function as an~\textit{input} to the algorithm, and assumed that given a sequence and a biophysical state, a protein will necessarily be found in the lowest-energy conformation.  However, to correctly model reality, we must dig deeper.  

\subsubsection{Free energy}\label{free_energy}

~Physically, we must define the~\textit{energy} of a conformation $c$ as a quantity proportional to $-T\ln P(c)$, where $P(c)$ is the probability of finding the molecule in conformation $c$ and $T$ is the temperature.  Without loss of generality, we will choose a proportionality constant $R$ (this defines units for the energy); $R$ is the universal gas constant.  Since different biophysical states are ultimately just different regions of conformational space, this notion of energy suffices to perform any single- or multi-state design: we simply wish to maximize the probability of the molecule being in the state we desire.  The probability of a biophysical state $s$ is the sum (or integral) of the probabilities of each of its conformations $c\in C(s)$, and is thus proportional to the partition function $q_s$, where
\begin{gather}
q_s = \sum_{c\in C(s)} \exp\left(-\frac{E(c)}{RT}\right).
\end{gather}
It is often useful to work not with the partition function directly, but with the~\textit{free energy} $G_s = -RT\ln q_s$ of the state. Then, we simply design to reduce the free energy of desired states and increase the free energy of undesired states.  Importantly, as the temperature goes to 0, $G_s$ becomes simply the energy of the state's lowest-energy conformation, and thus we arrive at the more approximate formulation of multistate design presented in Section~\ref{multistate_problem}.  But this approximation introduces error at nonzero temperature, and algorithms have been developed to actually use $G_s$ at physiological temperatures and thus account for the distribution of energies across conformational space.  

Computing the partition function is unfortunately \#P-hard, analogously to similar calculations in statistics.  However, the partition function can be efficiently approximated in practice for a particular sequence and biophysical state, while modeling continuous flexibility, using the $K^*$ algorithm~\cite{K*,CFTR,alg_smb_textbook}.  The $K^*$ algorithm builds on DEE/A* to model a thermodynamic ensemble of low-energy conformations for the bound and unbound biophysical states of a protein that the user wishes to design for binding.  Moreover, design based only on GMECs has been shown not to recapitulate sequences designed with $K^*$ that performed well empirically~\cite{CFTR}.  

More efficient algorithms have also been developed for this problem.  
The \bbks algorithm~\cite{BBK*} uses an A* tree with nodes from many sequences to compute the same top sequences as $K^*$, and thus provide the same guarantees of accuracy as $K^*$, in time sublinear in the number of sequences.  Thus \bbks~achieves high efficiency while approximating free energy with continuous flexibility.     

\subsubsection{Improved energy functions}\label{improved_efunc}
We have not yet addressed one very important question: how do we accurately estimate $E(c)$ for a conformation $c$?  The most commonly used energy functions in protein design~\cite[Table 12.1, page 103]{alg_smb_textbook}, like AMBER, EEF1, and the Rosetta energy function, make many approximations due to their prioritization of speed over accuracy.  More accurate energy functions based on induced electric multipoles, quantum chemistry, and Poisson-Boltzmann solvation theory are available, but they are expensive, and they violate a key assumption of the discrete pairwise model of protein design: they are not explicitly a sum of terms depending on at most 2, or indeed on any small number of residues' conformations.  

One approach to these problems is to use discrete rotamers and precompute pairwise energies by choosing a ``reference'' conformation, perturbing it by 1 or a few rotamers at each position, and using the differences in energy between the perturbed and reference conformations as 1-, 2-, and sometimes 3-body energies.  This approach yields relatively accurate energies for many systems, using either the Poisson-Boltzmann solvation model~\cite{PB_pairwise} or the AMOEBA forcefield (featuring induced multipoles)~\cite{A*_tuples} as the energy function.  

A second approach is to {\em learn} a representation of the energy suitable for protein design, from a training set that can be generated with any energy function.  This approach has the advantages of accommodating continuous flexibility and of not requiring all the 1- through 2- or 3-body perturbed conformations from the reference conformation to be physically realizable (this can be an issue in the case of backbone flexibility).  Two algorithms in the {\sc osprey}~\cite{OSPREY3} protein design software exploit this approach: the EPIC algorithm learns a polynomial approximation of the continuous energy surface within a voxel, and the LUTE algorithm~\cite{LUTE_RECOMB} directly learns a pairwise energy matrix (possibly augmented by triples) from sampled single-voxel minimized energies.  Both EPIC and LUTE have been shown to achieve small residuals, while calling the energy function just enough to obtain an accurate characterization of the energy costs of design decisions.  Thus, they greatly accelerate design using energies from quantum chemistry and Poisson-Boltzmann solvation~\cite{LUTE_RECOMB}.  

\section{``Exotic'' objective functions}\label{exotic_objfcn}

Not all protein design algorithms optimize energy with respect to sequence; we now review two other approaches.  

No matter how tightly a designed protein therapeutic binds its desired target, a strong reaction by the human immune system against this new protein may prevent it from remaining in the body for long, rendering it ineffective in the clinic.  The EpiSweep algorithm~\cite{EpiSweep} addresses this problem by finding sequences on the Pareto frontier between an \osprey-based~\cite{OSPREY_MIE,GrsA-LeuA,OSPREY3} stability design, and an objective function based on avoiding an immune reaction.  

It is also sometimes useful, even when optimizing binding, to search the space of known protein backbone conformations to find one that will place sidechains in a desired pose.  The RosettaMatch~\cite{RosettaMatch}, SEEDER~\cite{SEEDER}, and MASTER~\cite{MASTER} algorithms attack this problem.

\section{Protein design on graphics processing units}\label{gpu}
In the past decade, graphics processing unit (GPU) computation has transformed nearly every area of computational science, from molecular dynamics to computer vision to quantum chemistry.  Unlike early GPUs, which were difficult to use for applications other than graphics, today's GPUs are relatively easy to program using C-like languages like CUDA.  For suitably structured computations, GPUs can perform about 1000 times more floating-point operations per second per dollar spent on computational hardware.  

In the past few years, the computational tasks that are bottlenecks in protein design computation have been implemented for GPUs.  For the pairwise discrete model, the bottleneck is combinatorial optimization, and the gOSPREY software~\cite{gosprey} implements the A* algorithm on GPU hardware to accelerate this step.  For continuously flexible protein design, continuous energy minimization within a voxel is the bottleneck.  Thus, the \osprey software, which pioneered minimization-aware protein design, allows continuous energy minimization on GPUs as of its version 3.0~\cite{OSPREY3}.  Significant speedups (generally 1-2 orders of magnitude) were observed for designs.  This compares favorably with the previous flagship application of GPUs in computational structural biology, which is molecular dynamics (MD) simulations of proteins (temporal simulation of proteins using the classical mechanical potential defined by an energy function). 

GPUs can exploit two types of parallelism in order to accelerate the biomolecular energy computations central to MD and protein design: (a) processing different conformations of a protein in parallel, and (b) processing different parts of the molecule in parallel.  MD is better positioned to exploit (b) than protein design is, because MD evaluates energies for the entire molecule rather than merely the region around the mutations.  On the other hand, continuously flexible protein design can minimize energies for a huge number of conformations in parallel, while MD must proceed through different conformations (i.e., timesteps) in sequence.  This type (a) parallelism in protein design applies both to conformations enumerated in order of lower bound, as in iMinDEE~\cite{iMinDEE}, and to conformations sampled for the purpose of learning a discrete model of the continuously minimized energy, as in LUTE~\cite{LUTE_RECOMB}.  

Thus, the success of GPUs in accelerating MD computations and the favorable parallelizability of protein design compared to MD bode well for the prospect of very efficient continuously flexible protein design on GPUs, which is already quite impressive in \osprey 3.0~\cite{OSPREY3}.  

\section{Successful applications of computational protein design}\label{applications}
\begin{figure}
\includegraphics[width=3.2in]{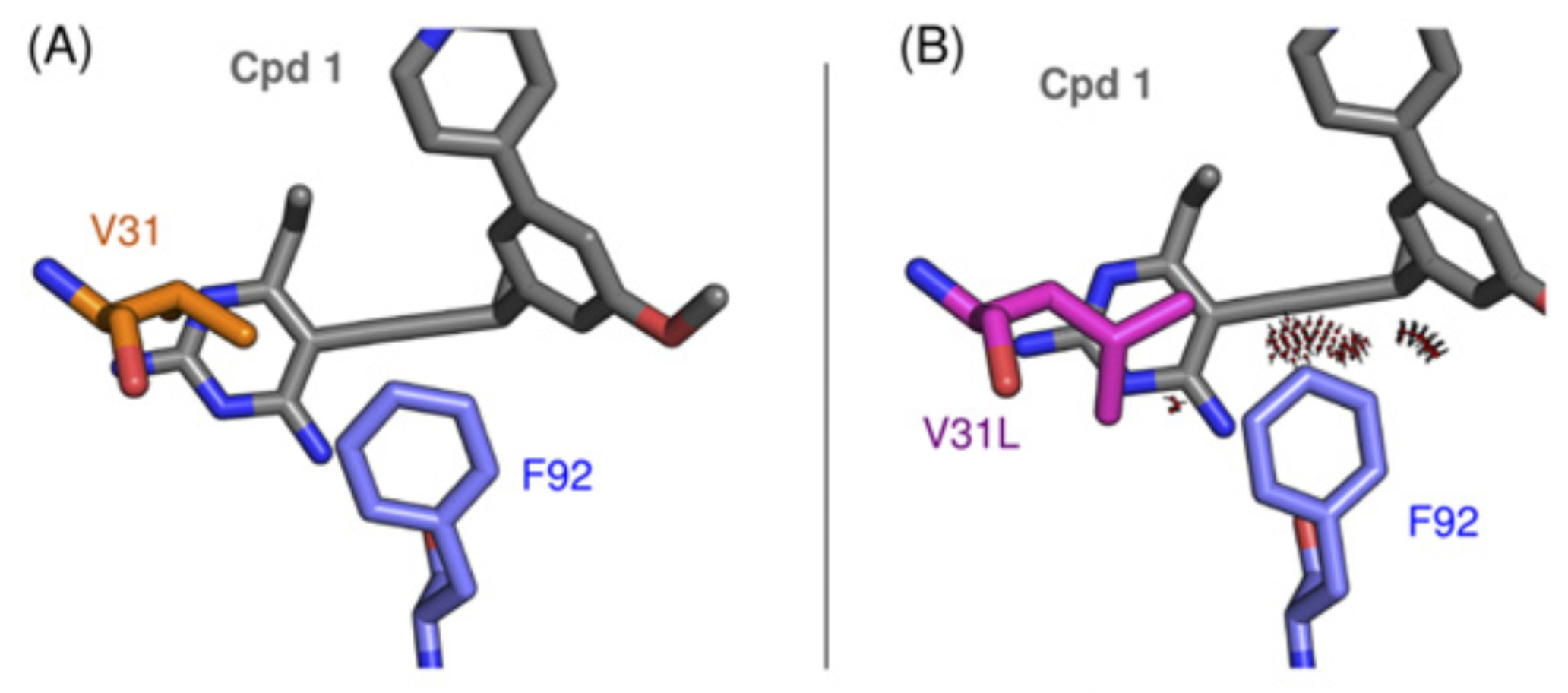}
\caption{Computational prediction of antibiotic resistance: (A) the bacterial (\textit{Staphyoloccus aureus}) enzyme dihydrofolate reductase binds a drug candidate (``Cpd 1'') tightly, inhibiting the enzyme's function, but (B) mutating position 31 of the enzyme from amino-acid type valine to leucine causes steric clashes that impeded binding, allowing the bacteria to resist the antibiotic.  This predicted resistance mutation was observed experimentally after being predicted by the $K^*$ algorithm as implemented in the {\sc osprey}~\cite{OSPREY3} software.   Figure adapted with permission from~\cite{DHFR-PNAS2}.  }
\label{resistance_prediction}
\end{figure}

Computational protein design algorithms have already produced many designs that perform well in experimental tests~\cite{alg_smb_textbook,cosb_design,OSPREY_MIE}.  Computationally designed enzymes have imparted completely new function to a protein~\cite{KempEliminase}, albeit with catalytic function still significantly below that of native enzymes, and have also exhibited a shift in substrate specificity from one ``molecular operand'' (input molecule) to another~\cite{GrsA-LeuA}.  Moreover, computational design algorithms can predict bacterial mutations in enzyme-coding genes that make the bacterial enzymes resistant to particular antibiotics (Fig.~\ref{resistance_prediction}), and these predictions have been confirmed both~\textit{in vitro}~\cite{DHFR-PNAS} and~\textit{in vivo}~\cite{DHFR-PNAS2}.  Computational protein design algorithms also excel at designing proteins with novel folds, an art form pioneered with the Top7 protein~\cite{Top7}.  

Finally, and perhaps most importantly, computationally designed proteins have shown promise in the design of therapeutics.  Using the techniques reviewed in this paper (in particular, the $K^*$ algorithm~\cite{VRC07_enhance} in \osprey~\cite{OSPREY3}), we collaborated with the NIH Vaccine Research Center to design a broadly neutralizing antibody against HIV with unprecedented breadth and potency (i.e., stronger activity against a broader range of HIV strains) that is now in clinical trials (Clinical Trial Identifier: NCT030151817). The \osprey/$K^*$ algorithm has also produced peptides that inhibit a protein involved in cystic fibrosis~\cite{CFTR}.  In addition to such direct design of therapeutics, computational prediction of resistance mutations to drug candidates~\cite{DHFR-PNAS,DHFR-PNAS2} will help combat resistance against new drugs (especially antibiotics) entering the clinic.  

\section{Conclusions}

Computational protein design has advanced significantly in the last decade.  Algorithms for the pairwise discrete approximation have matured, and significant progress is being made with improved biophysical models and for the design of clinically relevant proteins and peptides.  Proteins, especially antibodies, are attracting increasing attention from the pharmaceutical industry as drug candidates.  Protein design algorithms also have the potential to be transformative in the design of non-protein drugs, because unlike most drug design algorithms, they can search a large space of drug candidates in time sublinear in the the size of the space.  

To achieve protein design's full potential, it is necessary to further improve the accuracy of the biophysical model.  More accurate energy functions, improved modeling of protein-water interactions, and modeling of broader conformational spaces (both for search and for entropy computations) are likely to be important here.  As work continues on these important problems, the future of computational protein design looks bright.  

\paragraph{Acknowledgments}
We would like to thank Lydia Kavraki and Tom\'as Lozano-P\'erez for helpful discussions on the relationship between continuous and discrete algorithms; Nate Guerin, Jeff Martin, and Pablo Gainza for helpful comments on this manuscript; and all members of the Donald lab for many helpful discussions on protein design.  We also would like to thank all researchers in computational protein design for producing far more high-quality work than we have space to cover in this review, and we apologize to those whose work we were unable to include.  Finally, we would like to thank Toyota Technological Institute at Chicago (M.A.H.) and the NIH (grants R01 GM-78031 and R01 GM-118543 to B.R.D.) for funding.  

\bibliographystyle{plain}
\bibliography{../references}

\end{document}